\documentclass[%
 reprint,
 amsmath,amssymb,
 aps,
 prapplied,
superscriptaddress,
longbibliography
]{revtex4-2}

\usepackage{textcomp}

\usepackage{graphicx}
\usepackage{dcolumn}
\usepackage{bm}

\begin{document}

\title{Direct polariton-to-electron tunneling in quantum cascade detectors operating in the strong light-matter coupling regime}
\author{M. Lagr{\'e}e}\email{mathurin.lagree@3-5lab.fr}
\affiliation{III-V Lab, Campus Polytechnique, 1, Avenue Augustin Fresnel, RD 128, 91767 Palaiseau cedex, France}
\author{M. Jeannin}\thanks{These authors contributed equally}
\affiliation{Centre de Nanosciences et de Nanotechnologies (C2N),  CNRS UMR 9001, Universit{\'e} Paris-Saclay, 91120 Palaiseau, France}
\author{G. Quinchard}\thanks{These authors contributed equally}
\affiliation{III-V Lab, Campus Polytechnique, 1, Avenue Augustin Fresnel, RD 128, 91767 Palaiseau cedex, France}
\author{O. Ouznali}
\affiliation{Centre de Nanosciences et de Nanotechnologies (C2N),  CNRS UMR 9001, Universit{\'e} Paris-Saclay, 91120 Palaiseau, France}
\author{A. Evirgen}
\affiliation{III-V Lab, Campus Polytechnique, 1, Avenue Augustin Fresnel, RD 128, 91767 Palaiseau cedex, France}
\author{V. Trinit{\'e}}\email{virginie.trinite@3-5lab.fr}
\affiliation{III-V Lab, Campus Polytechnique, 1, Avenue Augustin Fresnel, RD 128, 91767 Palaiseau cedex, France}
\author{R. Colombelli}\email{raffaele.colombelli@u-psud.fr}
\affiliation{Centre de Nanosciences et de Nanotechnologies (C2N),  CNRS UMR 9001, Universit{\'e} Paris-Saclay, 91120 Palaiseau, France}
\author{A. Delga}
\affiliation{III-V Lab, Campus Polytechnique, 1, Avenue Augustin Fresnel, RD 128, 91767 Palaiseau cedex, France}

\begin{abstract}
Modern optoelectronic devices rely on cavity electrodynamics concepts for improved performances, embedding the medium hosting the electronic resonance in an optical cavity to enhance the light-matter coupling. This coupling rate is usually small compared to the electronic and photon energies. Despite several demonstrations, practical devices operating with much larger light-matter coupling strength, in the so-called strong light-matter coupling regime, are yet to be demonstrated as viable candidates. One of the main technological lock impeding their dissemination is the understanding of the carrier current extraction from strongly-coupled light-matter states. Here, we study this fundamental phenomenon in mid-infrared quantum cascade detectors (QCD) operating in the strong light-matter coupling regime. They operate around $\lambda$ = 10 $\mu$m with a minimum Rabi splitting of 9.3 meV. A simple model based on the usual description of transport in QCDs does not reproduce the polaritonic features in the photo-current spectra. On the contrary, a more refined approach, based on the semi-classical coupled modes theory, is capable to reproduce both optical and electrical spectra with excellent agreement. By correlating absorption/photo-response with the simulations, we demonstrate that - in this system - resonant tunneling from the polaritonic states is the main extraction mechanism. The dark intersubband states are not involved in the process, contrary to what happens in electrically injected polaritonic emitters.\\
\end{abstract}

\maketitle


\section{Introduction}

Polaritons, hybrid light-matter quasiparticles that arise when an electromagnetic field is strongly coupled to a matter (electronic) excitation, are an ideal playground for exploring  new physical phenomena and applications. Their matter part can be any polarization-carrying excitation, among which phonons \cite{Foteinopoulou2019Phonon}, excitons \cite{Weisbuch1992Observation}, molecular vibrations \cite{Shalabney2015Coherent}, cyclotron transitions \cite{Scalari2012Ultrastrong} or intersubband (ISB) transitions \cite{todorov2012intersubband, manceau2014mid}. Despite the large variety of polariton families, they share a common characteristics that motivates the continuous research in polariton physics: the light-matter coupling rate $\Omega$ can redefine the dominant energy or time scale in the system over those of the uncoupled constituents. Hence, they have been used as versatile platforms to test various fundamental phenomena \cite{Carusotto2013Quantum} such as Bose-Einstein condensation \cite{Kasprzak2006Bose}, topological effects \cite{Jacqmin2014Direct}, ultra-strong coupling \cite{Laurent2015Superradiant,Scalari2012Ultrastrong}, or the modification of energy landscapes in chemical reactions \cite{Herzog2019Strong}. Simultaneously to the constant progress in understanding polariton physics, they are pushed towards applications, with e.g. mid-infrared (MIR) optical polaritonic devices \cite{Pirotta2021Fast, jeannin2020unified} or as a new degree of freedom to alter the transport characteristics of electronic systems \cite{Orgiu2015Conductivity, Naudet2019Dark, Appugliese2021Breakdown}. 

One of the most widespread application target across the diverse polariton communities is the polaritonic emitter, producing coherent radiation with a higher efficiency than its counterpart operating in the weak light-matter coupling regime (e.g. LEDs or lasers). Coherent, laser-like emission from microcavity exciton polaritons condensates has been demonstrated in the visible part of the electromagnetic spectrum \cite{Fraser2016Physics}, while coherent mid-infrared thermal emitters have been realized \cite{Greffet2002Coherent, DeZoysa2012Conversion, Lu2020Narrowband} for applications that do not require laser light. Efforts are ongoing to develop MIR and THz lasers based on ISB polaritons \cite{ciuti2005quantum}, arising from the strong light-matter coupling between an ISB transition and a photonic micro-cavity mode, as they are predicted to have improved functionalities with respect to current technology \cite{de2008quantum, colombelli2015perspectives}.

However, efficient electrical injection into polaritonic devices is a major challenge, with few experimental demonstrations present in the literature \cite{sapienza2008electrically, Jouy2010Intersubband}. The reason is that most of the electrons injected from an electronic reservoir into a polaritonic system tunnel into dark states. These states do not couple to the electromagnetic field \cite{de2008quantum} and do not contribute to the emission process. 

To circumvent the injection problem and to get insightful information on the transport mechanisms between electronic and polaritonic states, the authors of Ref. \cite{vigneron2019quantum} studied the reverse phenomenon: the extraction of an electrical current from a polaritonic reservoir, i.e. a detection process. Indeed, previous reports of photo-detection in the strong coupling regime \cite{Dupont2003Vacuum, Sapienza2007Photovoltaic} brought little information on the electronic extraction process. The work in Ref. \cite{vigneron2019quantum} demonstrated photo-current extraction from microcavity polaritons into the continuum of electronic states using quantum well photodetectors and proposed an empirical model of the current extraction. Although this is an important step towards a better understanding of polaritonic devices, in the perspective of developing efficient light emitting devices, there is a strong need to study the \textit{resonant} current extraction from a polaritonic state into an electronic state, as well as to clarify the importance of dark states in the extraction process. Quantum cascade detectors (QCD)  \cite{gendron2004quantum,schwarz2017limit,delga2020quantum},  photo-voltaic counterparts of quantum well photo detectors, rely on the resonant extraction of electrons into a cascade of discrete energy levels, and thus are perfectly suitable to further elucidate the problem.  

In this context, we investigate MIR QCDs  ($\lambda = 10$~\textmu m) embedded in patch microcavities, operating at the onset of the strong light-matter coupling regime  ($2\Omega_\text{Rabi} = 9.3 $ meV). We observe the characteristic spectral features of this regime on both reflectivity and photo-current measurements.  
Using a semi-classical model based on the coupled mode theory (CMT) and by explicitly incorporating the spectral properties of the extraction cascade, we are able to reproduce quantitatively both reflectivity and photo response spectra. 

From this analysis, we can conclude that there is a direct resonant tunneling process between the polaritonic states and the electronic extractor state, and that dark states are not involved.
It is therefore possible to circumvent the problem of dark states when extracting an electrical current from a polaritonic reservoir.

\section{Device design and passive characterizations}\label{section2}
\subsection{Fabrication and measurements}
\begin{figure}[ht]
\centering\includegraphics[width=0.5\textwidth]{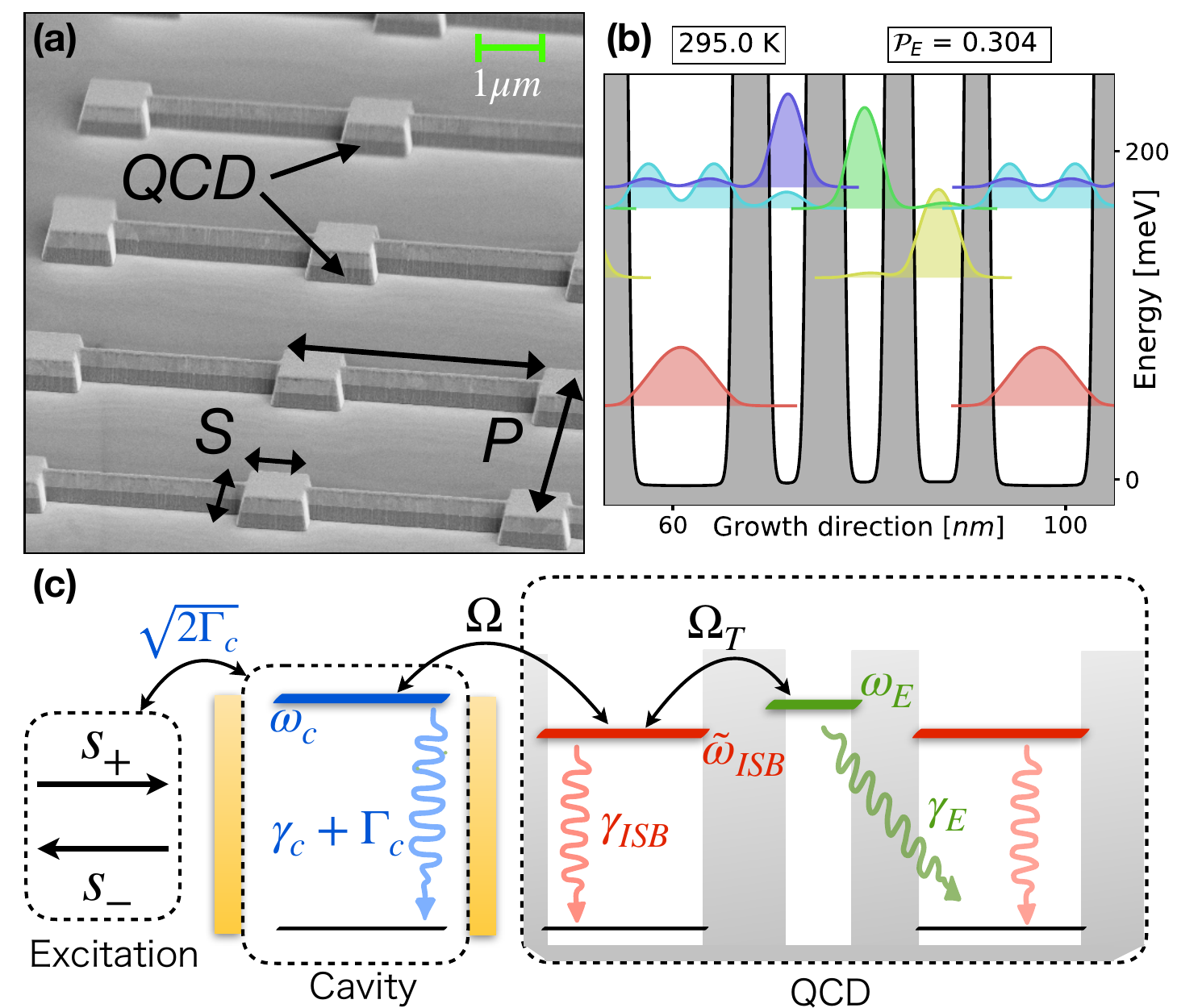}
\caption{\textbf{(a)} SEM image of a patch cavity-embedded QCD detector. $S$ is the patch lateral size and $P$ is the array period. Patches are electrically connected using gold wires deposited on a dielectric bridge layer. The active layers, the quantum cascade detectors (QCD), are embedded between gold layers (Au). \textbf{(b)} Bandstructure of the QCD (\underline{\textbf{11.0}}/3.6/\textbf{4.2}/3.8/\textbf{4.6}/2.8/\textbf{5.3}/2.5 nm, starting with the doped quantum well $n_\text{2D}=5e11$ cm$^{-2}$, InGaAs in bold, AlInAs normal), represented on one period ($T=295$K). \textbf{(c)} Schematic representation of the system: the incoming light excitation ($\omega)$ couples to the optical cavity ($\omega_c$), which is coupled to the intersubband transition ($\tilde{\omega}_\text{ISB}$). The ISB mode is itself coupled to an extractor mode ($\omega_E$), representing the electronic cascade. The different dissipation channels are represented ($\gamma_c$, $\Gamma_c$, $\gamma_\text{ISB}$ and $\gamma_E$). }\label{Fig:figure1}
\end{figure}

We first investigate the optical properties of patch antenna arrays embedding the quantum cascade detector (QCD) active region. The metal-metal patch resonators are obtained through waferbonding on a metallized Si substrate. They are processed into 1$\times$2~mm$^2$ matrices of period $P$ and patch dimensions $S$ with different $(P,S)$ values (Fig.~\ref{Fig:figure1}(a)). Changing the period $P$ allows to optimize the optical coupling into the cavities, while the patch size $S$ tunes the resonance frequency. 
The active layers are grown by molecular beam epitaxy (MBE). They form a 5-period InGaAs/AlInAs QCD detector, surrounded by 50 nm, n-doped (Si, $n_\text{3D} = 6\times10^{17}$~cm$^{-3}$) InGaAs contacts. 
The active quantum wells, where the intersubband absorption takes place, are n-doped (Si, $n_{2D} = 5\times 10^{11}$~cm$^{-2}$). The QCD band-structure (Fig.~\ref{Fig:figure1}(b)) is designed using a numerical software (METIS) relying on a refined and self-consistent Schr{\"o}dinger-Poisson scheme \cite{trinite2011modelling,terazzi2012transport}. Computations account for electron wavevector dispersion, non-parabolicity effects as well as coherent tunneling. The electronic subbands are aligned such that undesirable diagonal transitions between levels (resulting in dark current leakages) are minimized, while photo-current extraction is maximized through the optimization of the cascade alignment and barrier thickness. We compute a $0.3$ extraction probability $\mathcal{P}_E$ at 295~K. A conceptual representation of the whole structure is displayed on Fig.~\ref{Fig:figure1}(c).\\

Two samples have been processed for reflectivity measurements: a sample with undoped active layers inside the patch resonators (the empty cavity set)  and a sample with the Si-doped quantum wells. In this experiment, the patch antennas are not electrically connected by wires. The measurements are carried out using a Fourier transform infrared spectrometer (FTIR, Bruker Vertex 70) using a pair of parabolic ($F=180$~mm) and elliptic mirrors ($F=282/42$~mm) and a DTGS MIR detector. The reflectivity of the sample is normalized by the reflectivity spectrum of a Au mirror reference. All the measurements are performed at 295~K with a 13° incidence angle and the light is $p$-polarized. For both samples (doped and undoped), the reflectivity $\mathcal{R}$ is measured for different couples ($P,S$) of patch dimensions. 

\begin{figure*}[ht!]
\centering\includegraphics[width=0.8\textwidth]{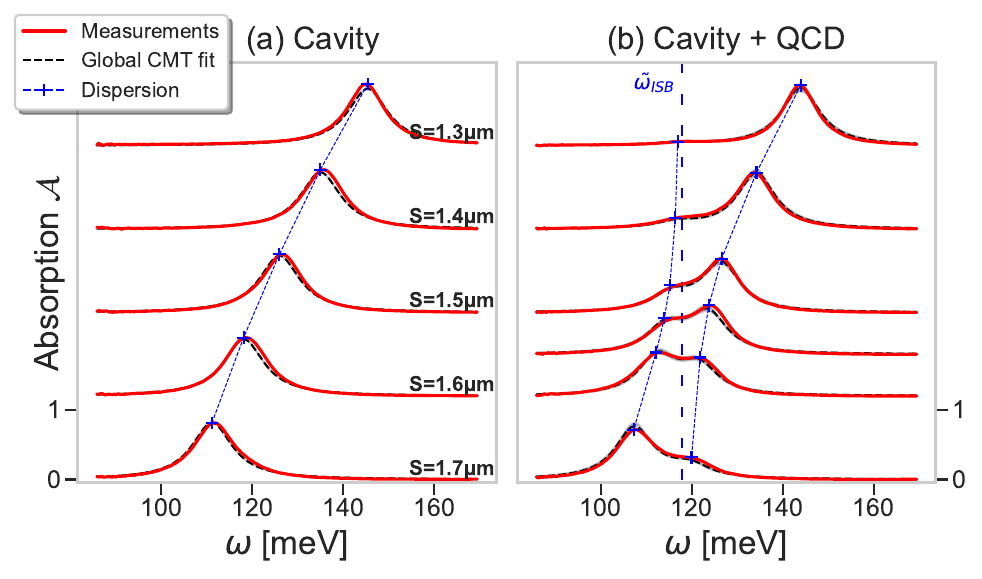}
\caption{Reflectivity measurements (continuous lines) and CMT global fit (dashed lines), represented in the form of absorption $\mathcal{A} = 1-\mathcal{R}$, for different patch sizes $S$ and fixed period $P=5$~\textmu m. The scale used is the same for all spectra and offsets are added for visibility. \textbf{(a)} Empty cavities (undoped active layers). Cavity dispersion (blue crosses) is computed using Eq.\eqref{Eq:dispersion} ($n_\text{eff}=3.28$) \textbf{(b)} Doped QCDs ($n_\text{2D} =5e11$~cm$^{-2}$) embedded inside the patch cavities. Polaritonic dispersion is computed using the Hamiltonian $\mathcal{H}$ in Eq.\eqref{Eq:CMTterms} ($n_\text{eff}=3.33$, $\omega_\text{ISB} = 115$~meV, $\omega_P=25$~meV). Additional measurements and fits are available in Supplemental Material (Fig.~S1). Although extremely low, propagated errors from the fit are represented on the spectra with the thickness of the grey curves. } 
\label{Fig:figure2}
\end{figure*}

We show in Fig.~\ref{Fig:figure2} the experimental absorption  $\mathcal{A} = 1-\mathcal{R}$ determined from the array reflectivity $\mathcal{R}$ (continuous lines) for the undoped sample (Fig. \ref{Fig:figure2}(a)) or the 
doped sample (Fig. \ref{Fig:figure2}(b)). The undoped sample exhibits a single 
absorption peak corresponding to the cavity mode only, that blue shifts in frequency for decreasing patch size $S$. 
The doped sample exhibits two resonances for each cavity size, 
which anticross around $\omega \approx120~$meV for a cavity size $S\approx1.55$~\textmu m. A minimal splitting of $\approx 9$~meV is estimated: this characteristic anticrossing around the ISB transition energy is a signature of the strong coupling regime. We detail below a quantitative assessment of the polaritonic features in the absorption spectra.

\subsection{Two-resonator coupled-mode-theory (CMT)}
To model the optical response of the system, we use a semi-classical formalism, the coupled modes theory (CMT). It describes in a phenomenological way the coupling between resonators and the dissipative interaction with their environment \cite{miller1954coupled,haus1984waves}. The CMT is a straightforward and powerful formalism that has been repeatedly used in the literature to successfully describes the optical response of intersubband devices embedded in metal-insulator-metal (MIM) resonators \cite{vigneron2019quantum,manceau2014mid,manceau2017immunity,jeannin2020absorption,jeannin2020unified}. Both of the ISB plasmonic mode and electromagnetic cavity mode are represented as coupled harmonic oscillators. The generic equation describing the evolution of the modes amplitude $\mathbf{a}$ interacting with one excitation port is \cite{zanotto2014perfect,fan2003temporal,suh2004temporal,zhao2019connection}:   
\begin{eqnarray}
\frac{\text{d}\mathbf{a}}{\text{d}t} = \left[i \mathbf{\mathcal{H}} - \left(\mathbf{\gamma + \Gamma }\right)\right] \mathbf{a} + K s_+ e^{i \omega t} \label{equationCMT}
\end{eqnarray} 
with: 
\begin{widetext}

\begin{eqnarray}
\mathcal{H} =  
\begin{bmatrix}
\omega_c & \Omega \\
\Omega & \tilde{\omega}_{\text{ISB}}
\end{bmatrix},  ~~~
\gamma + \mathbf{\Gamma} = \begin{bmatrix} 
\gamma_c + \Gamma_c & 0 \\
0 & \gamma_{\text{ISB}} 
\end{bmatrix}, ~~~
\mathbf{a} = \begin{bmatrix}
a_c \\ a_{\text{ISB}}
\end{bmatrix}, ~~~ 
K = \begin{bmatrix}
\sqrt{2 \Gamma_c} \\0
\end{bmatrix} \label{Eq:CMTterms}
\end{eqnarray}
\end{widetext}
The hermitian Hamiltonian $\mathcal{H}$ describes the unitary evolution of the system, $\mathbf{\Gamma}$ the dissipative interaction with the impinging light excitation (i.e. the radiative coupling), $\gamma$ the dissipative interaction with all the other external modes and $K$ the coupling between the system and the excitation. $|s_+|^2$ is the power of the impinging light excitation and $|a_i|^2$ is the energy stored in the $i$ resonator. The rotating wave approximation is applied, meaning that any contribution from antiresonant modes ($-\omega_c$ and $-\tilde{\omega}_{\text{ISB}}$) is neglected and that the system is restricted to two resonators. \\
  
The first resonator is the optical cavity, modeling the $\text{TM}_{01}$ electromagnetic mode confined in the patch antennas. The dispersion relation of the mode is directly related to the lateral size $S$ of the cavity \cite{todorov2010optical}:
\begin{eqnarray}
\omega_c = \frac{\pi c_0}{n_{\text{eff}} S} \label{Eq:dispersion}
\end{eqnarray}
The cavity frequency is thus entirely parameterized by $S$ and $n_{\text{eff}}$, the effective index of the cavity, which represents the effective medium composed of the semiconductor contacts and the undoped periodic structure embedded between the gold layers. $\gamma_c$ describes the mode losses which include all possible decay channels within the cavity (free-carrier losses and interaction with phonons mainly). \\

The second resonator models the intersubband absorption. The plasma-shifted energy $\tilde{\omega}_{\text{ISB}}$ is the re-normalized eigen energy of the ISB transition $\omega_\text{ISB}$, considering electron-electron interaction through the plasma energy $\omega_P$  \cite{todorov2012intersubband}: 
\begin{eqnarray}
\tilde{\omega}_{\text{ISB}}^2 = \omega_{\text{ISB}}^2 + \omega_P^2 \label{Eq:Rabi}
\end{eqnarray}
Note that this two-resonators CMT model implies that the ISB transition is homogeneously broadened, effectively assuming similar parabolic dispersion for both subbands. $\gamma_{\text{ISB}}$ describes the intersubband losses, mainly due to the LO phonon-scattering, interface roughness and alloy disorder. The cavity and intersubband resonators are coupled through the light-matter coupling constant $\Omega$ \cite{todorov2012intersubband}: 
\begin{eqnarray}
\Omega  &=& \frac{\omega_P}{2}\sqrt{f_w } 
\end{eqnarray}
with $f_w$ ($\approx 0.17$) the overlap factor between the cavity field and the doped active quantum wells. Since we expect an experimental deviation on the doping, we will use the plasma energy $\omega_P$ as a fitting parameter, in order to compute both ISB eigen energy $\tilde{\omega}_\text{ISB}$ and light-matter coupling $\Omega$. The Rabi splitting is defined as twice the coupling $2 \Omega$ and  corresponds to the minimal splitting between the polaritonic branches.

Incident light couples to the system through the cavity radiative losses $\Gamma_c$. The patch array can be assimilated to a 2D array of radiating slits, whose radiated power can be computed from interferometric considerations \cite{balanis2016antenna}. It allows to parametrize the radiative dissipative term $\Gamma_c$ as follows, with the coefficient $\alpha_c$: 
\begin{eqnarray}
\Gamma_c =  \frac{\alpha_c}{P^2}\label{Eq:cavityBroadening}
\end{eqnarray}
with $P$ the array period (Fig.~\ref{Fig:figure1}(a)). Adjusting $P$ allows in principle to reach the patch cavity critical coupling ($\gamma_c = \Gamma_c$) where all incident energy is dissipated in the system . 

The ISB mode does not couple directly to the incident light, it does so \textit{via} the cavity mode, such that $\Gamma_\text{ISB} \ll \Gamma_{c}$. However, before fully neglecting the direct radiative coupling of the ISB mode, we have to consider also the cross terms $\sqrt{\Gamma_\text{ISB}\Gamma_{c}}$ describing the radiative coupling of the modes with each other via the excitation channel \cite{suh2004temporal,le2017nanofiber}, as they can have considerable effects on the line-shapes in certain cases. At our doping levels~\cite{Ciuti-input-output}, the condition  $\Gamma_\text{ISB} \Gamma_{c} \ll\Omega^2$ is verified, so $\Gamma_\text{ISB}$ is indeed neglected. \\

In forced oscillation mode, the system of equations \eqref{equationCMT} is solved and the total absorption $\mathcal{A}$ is obtained considering the power dissipated through non-radiative processes inside the two resonators:
\begin{eqnarray}
\mathcal{A} = \mathcal{A}_c + \mathcal{A}_\text{ISB} = 2\gamma_c\frac{ |a_c|^2}{|s_+|^2} + 2\gamma_\text{ISB}\frac{ |a_\text{ISB}|^2}{|s_+|^2} \label{Eq:totalAbs}
\end{eqnarray}
where $\mathcal{A}_c$ and $\mathcal{A}_\text{ISB}$ are the cavity and ISB contributions to absorption. The absorption of the undoped sample (empty cavities) is computed using Eq.\eqref{equationCMT} and Eq.\eqref{Eq:totalAbs}, removing the coupling to the ISB mode ($\Omega =0$). It leads to the usual analytical expression of a Lorentzian absorption:
\begin{eqnarray}
\mathcal{A}_\text{Empty cavity} =  \frac{4 \gamma_c \Gamma_c}{(\Gamma_c + \gamma_c)^2 + (\omega - \omega_c)^2}\label{Eq:absorptionCavity}
\end{eqnarray}
Total absorption for the doped samples are computed using Eq.(\ref{Eq:totalAbs}) and numerically fitted to the experimental data \cite{lmfit_2021}. A global fit is performed using a unique set of parameters for the whole data set formed by each couple $(P,S)$, such that our problem is constrained to the maximal extent: the fitting parameters are ($n_\text{eff},\gamma_c,\alpha_c$) for the empty cavities, ($n_\text{eff},\gamma_c,\alpha_c,\omega_\text{ISB},\omega_P,\gamma_\text{ISB}$) for the doped samples. Results are presented in Fig.~\ref{Fig:figure2} (dashed black lines). More ($P,S$) couples are displayed in Fig.~S1 of Supplemental Material. \\

For both undoped (Fig.~\ref{Fig:figure2}(a)) and doped (Fig.~\ref{Fig:figure2}(b)) samples, we obtain an excellent agreement between the experimental data and the fit for the whole ($P,S$) dataset. Both cavity dispersion Eq.\eqref{Eq:dispersion} and $P$ dependency of the cavity broadening Eq.\eqref{Eq:cavityBroadening} are recovered. All along the anticrossing, the amplitudes of the polaritonic branches are well reproduced.

\begin{table*}[ht!]
\centering
\begin{tabular}{ |c|c||c||c|}

 \hline
  & \textbf{a.} Reflectivity fit :  bare cavity & \textbf{b.} Reflectivity fit :  $n_\text{2D}=5e11$cm$^{-2}$& \textbf{c.} Photo-current fit : $n_\text{2D}=5e11$cm$^{-2}$  \\
 \hline
 $n_{\text{eff}}$&$ 3.283 \pm 0.001$  & $3.333 \pm 0.001$ &  $3.219 \pm 0.001 $  \\
 $\gamma_c $ \small{(meV)}& $3.8 \pm 0.1$ & $3.5 \pm 0.01$   & $3.4 \pm 0.01$  \\
$\alpha_c $  \small{(meV.\textmu m$^2$)} &  $38.5 \pm 0.1$   &  $38.6\pm 0.1$  & $29.1 \pm 1.2$  \\
\hline
$\omega_{\text{ISB}}$ \small{(meV)} &- & $115.0 \pm 0.1$  & $115.1 \pm 0.1$  \\
$ \gamma_{\text{ISB}}$ \small{(meV)} & -& $5.2 \pm 0.1 $   & $3.3 \pm 0.1 $ \\
$\omega_P$ \small{(meV)}  &- & $25.6\pm 0.1$ & $29.2 \pm 0.1$\\
\hline
$\omega_E$ \small{(meV)} & -&- & $125.2 \pm 0.1 $ \\
$\gamma_E$ \small{(meV)} & -& -& $11.5 \pm 0.1$\\
$\Omega_T$ \small{(meV)} & -& -& $3.7 \pm 0.1$\\
\hline
\end{tabular}\\
 \caption{Parameters returned by global CMT fits of: \textbf{(a)} reflectivity of the undoped samples (Eq.\eqref{Eq:absorptionCavity}, $T=300$~K). \textbf{(b)} reflectivity of the doped samples (Eq.\eqref{Eq:totalAbs}, $T=300$~K.) \textbf{(c)} photo-current of doped samples, electrically connected (Eq.\eqref{Eq:photoCurrent}, $T=78$~K.) }
 \label{Table:fitParameters}
\end{table*}

Fit parameters are summarised in Table \ref{Table:fitParameters}. Cavity parameters $n_\text{eff}$, $\gamma_c$ and $\alpha_c$ are coherent between the doped and undoped samples. We obtain a plasma frequency of $26$ meV, which converts into a $2\Omega=9.3$~meV Rabi splitting, according to Eq. \eqref{Eq:Rabi}. Polaritonic broadenings are estimated to be $\frac{\gamma_c+\Gamma_c+\gamma_\text{ISB}}{2} \approx 5$~meV.  
Although we are only at the onset of the strong coupling regime, we are able to distinguish the two polaritonic contributions to the absorption spectra and estimate the Rabi splitting. The equivalent doping can therefore be computed using the plasma frequency expression:

\begin{eqnarray}
\omega_P^2 = \frac{e^2 \Delta n_{1\small{\rightarrow}2}f_{1\small{\rightarrow}2}}{\epsilon \epsilon_0 m^* L_\text{eff} } \approx \frac{e^2 n_\text{2D}f_{1\small{\rightarrow}2}}{\epsilon \epsilon_0 m^* L_\text{QW} } 
\end{eqnarray}
where $ \Delta n_{1\small{\rightarrow}2}$ is the electronic surface density difference between the lower and upper subbands of the optical quantum well, $f_{1\small{\rightarrow}2}$ is the oscillator strength of the ISB transition  ($= 0.84$, computed using METIS), $m^*$ is the effective electron mass and $L_\text{eff}$ is an effective length corresponding to the spatial extension of the ISB currents, that can be usually approximated by $L_\text{QW}$. We also neglect the electronic density thermally promoted to the excited state.
We find a $n_\text{2D}=3.7e11$ cm$^{-2}$ equivalent doping, $35\%$ lower than the expected one.
The ISB energy is estimated around $\omega_\text{ISB}=115$~meV, even though it was designed to be 120~meV. This experimental red-shift is the consequence of the non-parabolicity of the subbands, unaccounted for in the CMT. It can be computed using the METIS software : at 300K, we find that the absorption peak is indeed red-shifted toward 116 meV.

Polaritonic dispersion is computed via the diagonalization of the Hamiltonian $\mathcal{H}$ of Eq.\eqref{Eq:CMTterms} and superimposed on the spectra (blue dotted lines Fig.~\ref{Fig:figure2}(b)). The absorption peaks closely follow the dispersion and $\tilde{\omega}_\text{ISB}$ is indeed an asympotic value of the polaritonic branches. 
The fit parameters present little correlation between one another, which propagates into low errors on the spectra (see Section I of Supplemental Material for the detailed computations of the errors). The weak experimental/theoretical disagreements are rather related to our over-simplified model and to the experimental dispersion (experimental uncertainties on $S$, $P$, $n_{\text{2D}}$, $\omega_\text{ISB}$ mainly).

\section{Photo-current measurements and comparison with modeling}\label{section3}
\subsection{Measurements}
We are now interested in the photo-detection response of the devices operating in strong coupling. A set of smaller, 8$\times$8 ($\approx50\times50$~\textmu m$^2$) patch antenna arrays is processed. In this experiment, patches are connected through 250~nm thin metallic wires realized by electron-beam lithography (Fig. \ref{Fig:figure1}(a)). The samples are placed inside a cryostat ($T=78$K) at zero voltage applied bias. 
(It is necessary to operate in such experimental conditions: as a matter of fact, we observed that the application of a (small) bias leads to a drastic decrease in the SNR (signal to noise ratio) that hampers the photo-current measurement). 

Samples are illuminated by the FTIR globar source, and the signal is acquired in rapid scan mode. 
Light is $p$-polarized perpendicular to the wires, in order to excite the mode that is not perturbed by the wires~\cite{hakl01}.
The generated photo-current is amplified using a low noise trans-impedance amplifier.

\begin{figure*}[ht!]
\centering\includegraphics[width=0.8\textwidth]{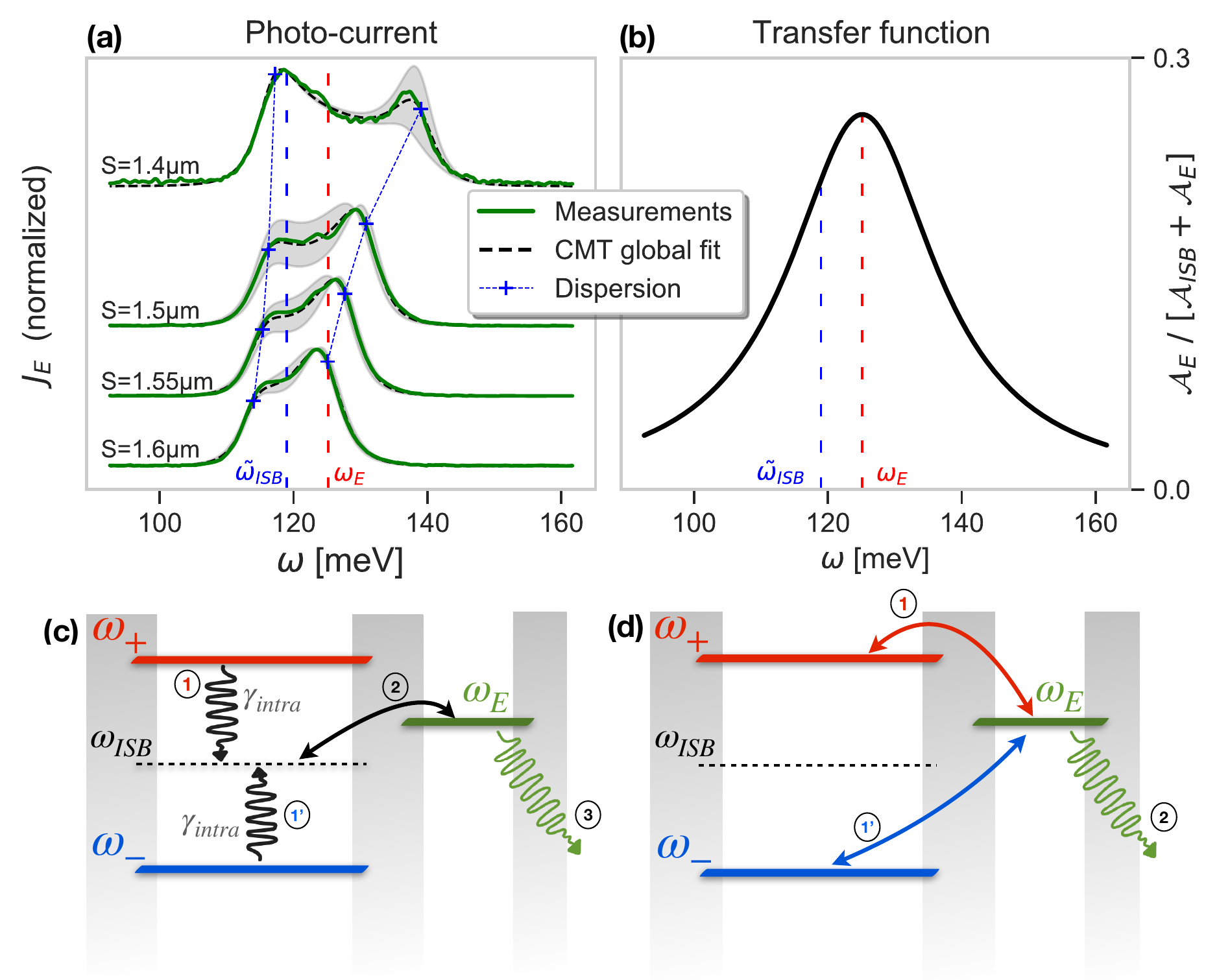}
\caption{ \textbf{(a)} Normalized photo-current measurements (continuous lines) and three resonators CMT global fit (dashed lines), for devices with period $P=7$~\textmu m and different patch sizes $S$. Offsets are added for visibility. The grey areas represent the fit errors propagated on the spectra (see Supplemental Material section I for detailed computation). 
The polaritonic dispersion (blue crosses) is computed considering only the cavity-ISB interaction in the Hamiltonian $\mathcal{H}$ of Eq.\eqref{Eq:3CMTterms}, unperturbed by the QCD extractor ($n_\text{eff}=3.23$, $\omega_\text{ISB} = 115$~meV, $\omega_P=29$~meV). 
\textbf{(b)} Computed transfer function between the extracted power ($\mathcal{A}_E$) and the total power dissipated inside the QCD ($\mathcal{A}_E + \mathcal{A}_\text{ISB}$). 
Lower panels: possible extraction schemes for a QCD in the strong light-matter coupling regime. 
\textbf{(c)} Dark-states mediated extraction: polaritonic excitations ($\omega_\pm)$ relax into the dark states ($\omega_\text{ISB})$, and they are subsequently extracted through the electronic cascade.
\textbf{(d)} Resonant tunneling through the polaritonic states: polaritonic excitations directly tunnel from the polaritonic reservoir into the electronic cascade. 
In this case the dark states are not involved in the process.}
\label{Fig:figure3}
\end{figure*}

Measurements on devices with $P=7$~\textmu m are displayed in Fig.~\ref{Fig:figure3} (continuous lines), and all the photocurrent spectra are normalized to unity. 
Additional measurements on devices with different ($P,S$) values are presented in Supplemental Material (Fig.~S2). 
The photocurrent spectra exhibit two anticrossing peaks, the spectral signature of the strong-light matter coupling regime. 
Equal peak amplitudes occur between $S=1.5$~\textmu m and $S=1.4$~\textmu m, for $\omega \approx 125$~meV. This does not correspond to the minimal splitting between the two polaritonic branches (around $S=1.6$~\textmu m). 
Far from this minimal splitting (e.g. $S=1.4$~\textmu m), the relative amplitude of the photo-current peaks appear to follow the opposite trend than the reflectivity measurements: the upper polaritonic mode has a smaller amplitude than the lower polaritonic one.
In the follow-up, we will show that these relatively small differences permit to infer important informations on the electronic transport.

Note: Although the detector samples stem from the same wafer as the ones used for reflectivity measurements in Section 2 (the active region is the same), they are not identical. 
They have been fabricated separately, and the patch cavities are not connected by wires in the samples used for reflectivity. 
For this reason, we can not make a direct one-to-one comparison between samples with the same nominal parameters.

\subsection{Modeling Strategy}
In the following section, we will attempt to model and reproduce these photo-current spectra, and we will show that their peculiar features permit to extract informations on the tunneling process between the polaritonic and electronic reservoirs in this system.

We will develop and apply two different photo-current models for QCDs operating in the strong light-matter coupling regime. 
The two approaches rely on different assumptions and model two different possible extraction schemes, as displayed in Figure~\ref{Fig:figure3}(c) and (d). 

In the first scheme (panel c), photo-excited polaritons \textit{first} decay into dark states, and then are extracted into the electronic reservoir, i.e. the QCD extractor. This is essentially the reverse of the process that has traditionally hampered polaritonic emitters.\\
The second scheme (panel d), on the other hand, does not involve the dark states: polaritons directly tunnel into the electronic reservoir.
Our goal is to understand if the experimental results can discriminate between the two processes and tell us which extraction scheme is at work in the studied devices.

\subsection{Dark-states mediated extraction: two resonators CMT \label{section:2CMT}}

In the weak light-matter coupling regime, the usual description of transport inside quantum cascade detectors and ISB devices in general relies on the hypothesis that intrasubband processes ($\gamma_{\text{intra}}$, scattering \textit{inside} the subbands) are significantly faster than intersubband processes \cite{koeniguer2006electronic,buffaz2010role}, and are responsible for the subbands thermalization. 
This allows a markovian and sequential vision of transport: after an intersubband hop, electrons are first thermalized inside the subband before the next hop, following the Fermi-Dirac distribution. Memory of the previous hop is lost, and hops are independent. The subbands in this picture are simple electronic reservoirs exchanging excitations. Electronic transport is treated with perturbation theory and semi-classical rate equations \cite{trinite2011modelling,delga2012master,saha2016rate}. 
In this framework, the photo-current $\mathcal{J}_E$ (at zero bias) is proportional to the ISB absorption \cite{delga2020quantum}: 
 \begin{eqnarray}
 \mathcal{J}_E(\omega) \propto \mathcal{P}_E \mathcal{A}_\text{ISB}(\omega) \label{Eq:sequential}
 \end{eqnarray}
The scalar transfer function $\mathcal{P}_E$ is the extraction probability, i.e. the probability that an electron promoted to the excited state of the ISB transition escapes to the ground state of the next adjacent period, through the extraction cascade. 
In the weak excitation regime, $\mathcal{P}_E$ is independent of the impinging light frequency $\omega$, because carriers always hop from the same state, regardless of the excitation frequency.
In this scenario, the generated photo-current has the same spectral shape as the ISB contribution to the absorption.

In the strong light-matter coupling regime instead, the Rabi oscillations $\Omega$ between the ISB mode and the cavity mode are faster than the dissipative and decoherent processes: we are able to observe the two characteristic peaks of the polaritonic modes in the reflectivity spectra. 
However, in the standard description of transport in QCDs, the tunnel coupling between the upper optical level and the first cascade level, represented by $\Omega_T$, is slower than the intrasubband processes $\gamma_\text{intra}$.
As a consequence, the usual sequential interpretation of the extraction process, represented in Figure \ref{Fig:figure3}(c), applies: polaritonic excitations are first optically pumped at frequencies $\omega_\pm$. 
Because of the fast intrasubband decoherence, the excited polaritons {\it first} collapse into the ISB dark modes $\omega_\text{ISB}$, {\it then} they are exactred - as electronic excitations - as they would be in a QCD operating in the weak coupling regime.
Such sequential process obviously implies a flat transfer function: a similar spectral shape between the ISB absorption $\mathcal{A}_\text{ISB}(\omega)$ and the photo-current $\mathcal{J}_E$ is expected in this scenario.

This dark-states mediated extraction scheme can be faithfully modeled by the two-resonator CMT model developed in Eq.\eqref{Eq:CMTterms}. Using a frequency-independent transfer function $\mathcal{P}_E$, the normalized photo-current spectra can be fitted using only the normalized ISB component $\mathcal{A}_\text{ISB}$ of the total absorption Eq.\eqref{Eq:totalAbs}. We have performed a global fit of the photo-current measurements using  $n_\text{eff}$, $\gamma_c$, $\alpha_c$, $\omega_\text{ISB}$, $\omega_P$, $\gamma_\text{ISB}$ as parameters, and the results are presented in Supplemental Material, Fig.~S3. 
The general quality of the fit is poor especially close to the anticrossing region. The spectral linewidths and the peak positions are not well reproduced. The computed spectra are also unable to reproduce the relative peak amplitudes of the polaritonic branches: for $S=1.5,~1.55$ and $1.6$~\textmu m, the computed relative peaks amplitudes are reversed with respect to the measurements.

These results suggest that the photo-current measurements display singular features that can not be reproduced by a model relying on a frequency-independent transfer function $\mathcal{P}_E$. 
This is equivalent to say that the dark-states mediated extraction, schematically represented in Fig.~\ref{Fig:figure3}(c), is probably not the process that enables operation of the QCDs operating in strong coupling.

\subsection{Resonant extraction: three resonator CMT \label{section:3CMT}}

We have therefore developed a new photo-current model that explicitly describes the extraction process and is able to model a resonant extraction mechanism, as sketched in Fig.~\ref{Fig:figure3}(d).

To this scope, we introduce a new mode in the CMT model, the extraction mode $\omega_E$. It effectively models the electronic cascade. 
A schematic representation of the CMT implementation of the  complete system - i.e. optical cavity, ISB transition and extraction - is displayed on Fig. \ref{Fig:figure1}(c). 
The extraction mode is coupled to the ISB mode through a tunnel coupling constant $\Omega_T$, and it dissipates electrons into the next period at a rate $\gamma_E$. 
Obviously, it can not directly couple to the incident light ($\Gamma_E = 0)$ . 
Equation \ref{Eq:CMTterms} is modified introducing a third resonator of amplitude $a_E$:
\begin{widetext}
\begin{eqnarray}
\mathcal{H} =  
\begin{bmatrix}
\omega_c & \Omega & 0 \\
\Omega & \tilde{\omega}_{\text{ISB}} & \Omega_T \\
0 &\Omega_T & \omega_E
\end{bmatrix},  ~~~
\gamma + \mathbf{\Gamma} = \begin{bmatrix} 
\gamma_c + \Gamma_c & 0 & 0 \\
 0 & \gamma_{\text{ISB}} & 0 \\
 0 & 0 & \gamma_E
\end{bmatrix}, ~~~
\mathbf{a} = \begin{bmatrix}
a_c \\ a_{\text{ISB}} \\ a_E
\end{bmatrix}, ~~~ K = \begin{bmatrix}
\sqrt{2 \Gamma_c} \\ 0 \\ 0
\end{bmatrix} \label{Eq:3CMTterms}
\end{eqnarray}
\end{widetext}
The system total absorption $\mathcal{A}$ is the sum of the power dissipated into the different decay channels, normalized by the incoming power $|s_+|^2$:
\begin{eqnarray}
\mathcal{A}_\text{TOT} &=& \mathcal{A}_c + \mathcal{A}_\text{ISB} + \mathcal{A}_E  \\ &=&  2\gamma_c\frac{ |a_c|^2}{|s_+|^2} + 2\gamma_\text{ISB}\frac{ |a_\text{ISB}|^2}{|s_+|^2} + 2\gamma_E\frac{ |a_E|^2}{|s_+|^2} \label{Eq:totalAbs3CMT}
\end{eqnarray}
The net photo-current $\mathcal{J}_E$ is defined as the current under illumination subtracted from the dark current. Since the measurements are performed at zero bias, there is no dark current and the photo-current  $\mathcal{J}_E$ is proportional to the power dissipated from a period to the next adjacent period. 
This is exactly the power  dissipated by the new extraction mode:
\begin{eqnarray}
\mathcal{J}_E \propto  2 \gamma_E |a_E|^2 \label{Eq:photoCurrent} 
\end{eqnarray}
In this theoretical framework, the transfer function between the photo-current and the power dissipated inside the QCD is no longer flat and independent of the impinging light frequency $\omega$: it has a Lorentzian shape, centered around the extractor mode frequency $\omega_E$ (Figure \ref{Fig:figure3}(b)). 
It is defined as $\text{TF}(\omega)$: 

\begin{eqnarray}
\text{TF}(\omega) = \frac{\mathcal{A}_E}{\mathcal{A}_\text{ISB} + \mathcal{A}_E}\label{Eq:TF}
\end{eqnarray}

A frequency dependent transfer function was used previously in the literature. Vigneron et al.~\cite{vigneron2019quantum} introduced it to model the transport mechanisms underlying the tunnel coupling process in a QWIP detector. 
Earlier, Sapienza et al.~\cite{sapienza2008electrically} introduced a phenomenological Gaussian transfer function between the electroluminescence and the absorption spectra.
One of the novelties of this work is that the Lorentzian transfer function directly arises from the model and it is not introduced as a phenomenological parameter.

To test the model against the experiments, we performed a global fit of the experimental photo-current spectra, using equation \ref{Eq:photoCurrent} with the following fit parameters: $n_\text{eff}$, $\gamma_c$, $\alpha_c$, $\omega_\text{ISB}$, $\omega_P$, $\gamma_\text{ISB}$, $\omega_E$, $\gamma_E$, $\Omega_T$. 
The photo-current curves arising from the modeling are represented in Figure \ref{Fig:figure3}(a) as dashed lines for devices with $P=7$~\textmu m. 
Additional ($P,S$) couples are reported in Fig.~S2 in Supplemental Material. 
We obtain an excellent agreement between the experimental and computed spectra. We are now able to reproduce both peaks positions and linewidths. And - crucially - the model also correctly reproduces the relative peak amplitude of the polaritonic branches for all the cavity sizes $S$. 

The fitted parameters are reported in Table \ref{Table:fitParameters}(c). They are extremely consistent with the parameters returned by the reflectivity fit. The small variations observed between photo-current and reflectivity parameters are attributed to different measuring temperatures (lower losses $\gamma_c$ and $\gamma_\text{ISB}$ in particular for lower temperature) and the different technological processes between the two experiments.  The extraction mode energy $\omega_E$ is estimated to be about 125~meV, which is coherent with the predicted alignment between the upper optical level and extraction level, designed such that $\omega_E - \omega_\text{ISB}=10$~meV (Fig.~\ref{Fig:figure1}(b)). The tunnel coupling $\Omega_T=3.5$~meV is also equal to the value computed using our numerical software (METIS). Finally, the extraction rate $\gamma_E$ is estimated around 11 meV, significantly faster than the other dissipation rates $\gamma_\text{ISB}$ and $\gamma_c$, which is consistent with the idea that the cascade is an efficient extractor.

\subsection{Physical interpretation of the modeling results}
The three resonator CMT model reproducing well these  singular features of the photo-current spectra validates the process sketched in Fig.~\ref{Fig:figure3}(d): polaritons are extracted into the electronic reservoir \textit{via} a resonant process that does not involves the dark states.
It is important to detail a bit more the subtleties of the model.

Being a semi-classical formalism, CMT is only able to describe dissipative processes: damping dissipates both modes population $|a_i|^2$ and the coherence $a_i^*a_j$. 
The model is however unable to depict  pure decoherence effects, i.e. processes that only affect coherence, without dissipating the modes populations. Specifically in our system, the CMT is unable to describe the intrasubband dynamics of transport $\gamma_\text{intra}$, as it does not affect the subbands populations, only erasing the coherence between them (through thermalization). 
To explicitly integrate the decoherence processes, we would require a more rigorous formalism based on the density matrix and quantum master equations \cite{schlosshauer2019quantum}. 
The underlying hypothesis of this CMT model is therefore that the intrasubband processes are significantly slower than any other transport mechanisms and are neglected. 

Consequently, the three resonator CMT model describes the  extraction scheme represented in Figure \ref{Fig:figure3}(d). 
The excitations are first pumped into the polaritonic modes $\omega_\pm$. By assumption, the polaritonic excitations do not have the time to collapse into the dark ISB modes through the intrasubband processes $\gamma_\text{intra}$: the resonant tunnel extraction process $\Omega_T$ is faster. 
The polaritons are directly extracted from the polaritonic states into the electronic cascade, without any involvement of the dark ISB modes in the process.

The three resonator formulation of the CMT provides a compact and convenient model for a complicated system in a non-standard operating regime. It provides an intuitive representation of the system, and also a robust framework to analyse and interpret the experimental data.
In particular, the excellent agreement it provides between experiment and computed photo-current measurements, combined with the failure of the intrasubband dominated transport model of section \ref{section:2CMT}, permits to elucidate the nature of the polariton-to-electron transport in these devices.
The schematic representation of Figure \ref{Fig:figure3}(d) is revealed as an accurate description of the transport in a QCD operating in the strong light-matter coupling regime.

\section{Conclusion}
We have demonstrated quantum cascade detectors embedded in patch antenna resonators operating in the strong light-matter coupling regime ($2\Omega_R = 9.3$ meV). By correlating reflectivity and photo-response measurements, we investigate the resonant tunnel extraction of ISB microcavity polaritons into an electronic reservoir. The comparison with a specifically developed three-resonator CMT model allowed us to elucidate crucial details of the polaritonic transport.
We could show that - in this system - resonant tunneling from the polaritonic states is the main extraction mechanism : most of the polaritonic excitations are directly extracted into the electronic cascade before they collapse into the dark states due to the decoherent intrasubband scattering. The dark ISB states are not involved in the process, contrary to what happens in electrically injected polaritonic emitters.

This work demonstrates that quantum cascade detectors are key devices to study ISB polaritons. Experimentally, the natural follow-up of this work is the exploration of a wider range of ISB-cavity coupling rates. Higher doping is required ($n_\text{2D}\approx 10^{12} cm^{-2}$). On the theoretical side, it would be useful to explicitly incorporate the intrasubband dynamic $\gamma_\text{intra}$ using the density matrix formalism \cite{breuer2002theory,del2008quantum,schlosshauer2019quantum}. Finally, the demonstration that is is possible to engineer a resonant tunneling extraction of polaritons into an electronic reservoir, without involvement of the dark states, paves the way toward a better understanding of polaritonic transport, with the ultimate goal of implementing efficient electrical injection in polaritonic light-emitting devices~\cite{Colombelli_SST,sapienza2008electrically,colombelli2015perspectives}

\begin{acknowledgments}
We acknowledge S.Barbieri, J-F. Lampin and M. Hakl of the \textit{Institut d'Electronique, de Microélectronique et de Nanotechnologie} (IEMN)  for the electromagnetic simulation and design of the patch antenna arrays. \\
We acknowledge financial support from the French National Research Agency : project SOLID (ANR-19-CE24-0003-02), HISPANID (ANR-17-ASTR-0008-01) and IRENA (ANR-17-CE24-0016). \\
We acknowledge financial support  from the European Union FET-Open Grant MIR-BOSE (No. 737017).
\end{acknowledgments}

\bibliography{references}

\end{document}


\title{Supplemental materials : Direct polariton-to-electron tunneling in quantum cascade detectors operating in the strong light-matter coupling regime}

\author{M. Lagr{\'e}e$^{\dagger*}$, M. Jeannin$^*$, G. Quinchard$^\dagger$, O. Ouznali$^*$, A. Evirgen$^\dagger$, V. Trinit{\'e}$^\dagger$, R. Colombelli$^*$ and A. Delga$^\dagger$}
\affiliation{\vspace{10pt}$^\dagger$III-V Lab, Campus Polytechnique, 1, Avenue Augustin Fresnel, RD 128, 91767 Palaiseau cedex, France\\ \\
$^*$Centre de Nanosciences et de Nanotechnologies (C2N),  CNRS UMR 9001, Universit{\'e} Paris-Saclay, 91120 Palaiseau, France}
    
\maketitle
\section{Errors computation}
We explain how we are able to propagate the parameters errors, returned by the global fits of our experimental data, into errors on the computed spectra (both reflectivity and photo-current).\\ The fit procedure outputs the covariance matrix $M_P$ related to the fit parameters. The standard deviation vector $\sigma_P$, representing the errors made on the parameters, is given by : 
\begin{eqnarray}
\sigma_P = \sqrt{\text{diag}(M_P)}
\end{eqnarray}
Let $F$ be the function used by the fit and $J$ the jacobian of the function. In order to propagate the errors made on the fit parameters onto the function $F$, we first compute the covariance matrix of $F$, $M_F$ : 
\begin{eqnarray}
M_F = J M_P J^T
\end{eqnarray}
Such that the propagated errors of $F$ are given by the standard deviation vector $\sigma_F$ : 
\begin{eqnarray}
\sigma_F = \sqrt{\text{diag}(M_F)}
\end{eqnarray}

\section{Additional figures}
\begin{figure*}[ht!]
\centering\includegraphics[width=0.8\textwidth]{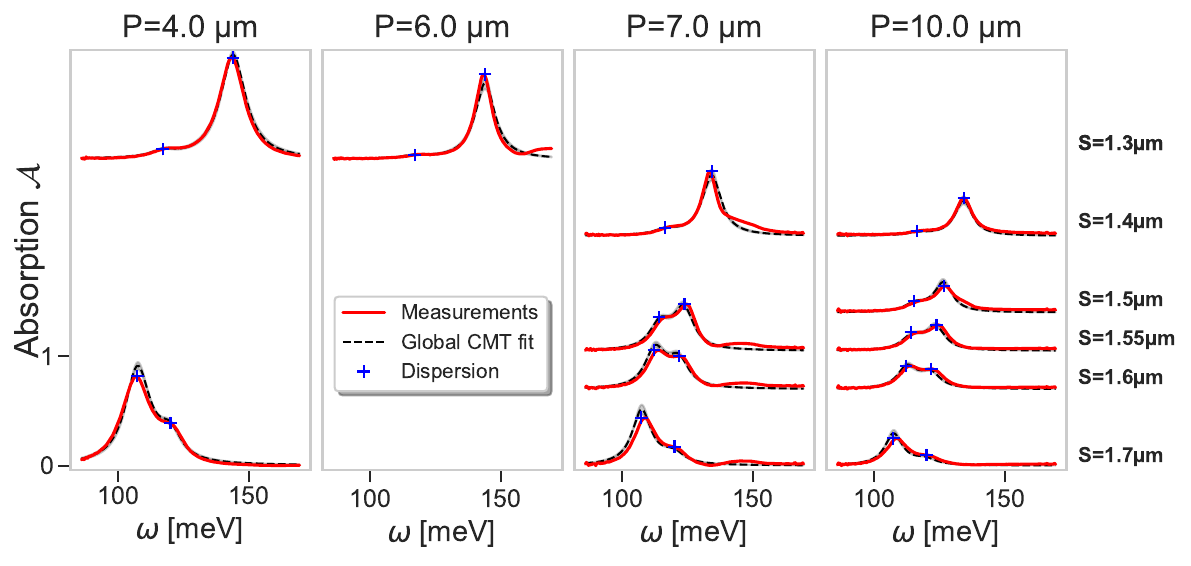} 
\caption{Additional reflectivity measurements (red continuous lines) and CMT global fit (black dashed lines), represented in the form of absorption $\mathcal{A} = 1-\mathcal{R}$, for different patch sizes $S$ and periods $P$, using doped QCDs ($n_\text{2D} =5e11$~cm$^{-2}$). The scale used is the same for all spectra and offsets are added for visibility.  Polaritonic dispersion (blue crosses) is computed using the Hamiltonian $\mathcal{H}$ in Eq.(2) of the main text ($n_\text{eff}=3.33$, $\omega_\text{ISB} = 115$~meV, $\omega_P=25$~meV). Although extremely low, propagated errors from the fit are represented on the spectra (grey areas). Note that the first diffraction peak is visible around 145 meV for $P=7$ \textmu m, and is mixed with the signal for $P=10$ \textmu . It is not included in the CMT model, but could be readily added should the need arise. }\label{supplementary1}
\end{figure*}
\begin{figure*}[ht!]
\centering\includegraphics[width=0.8\textwidth]{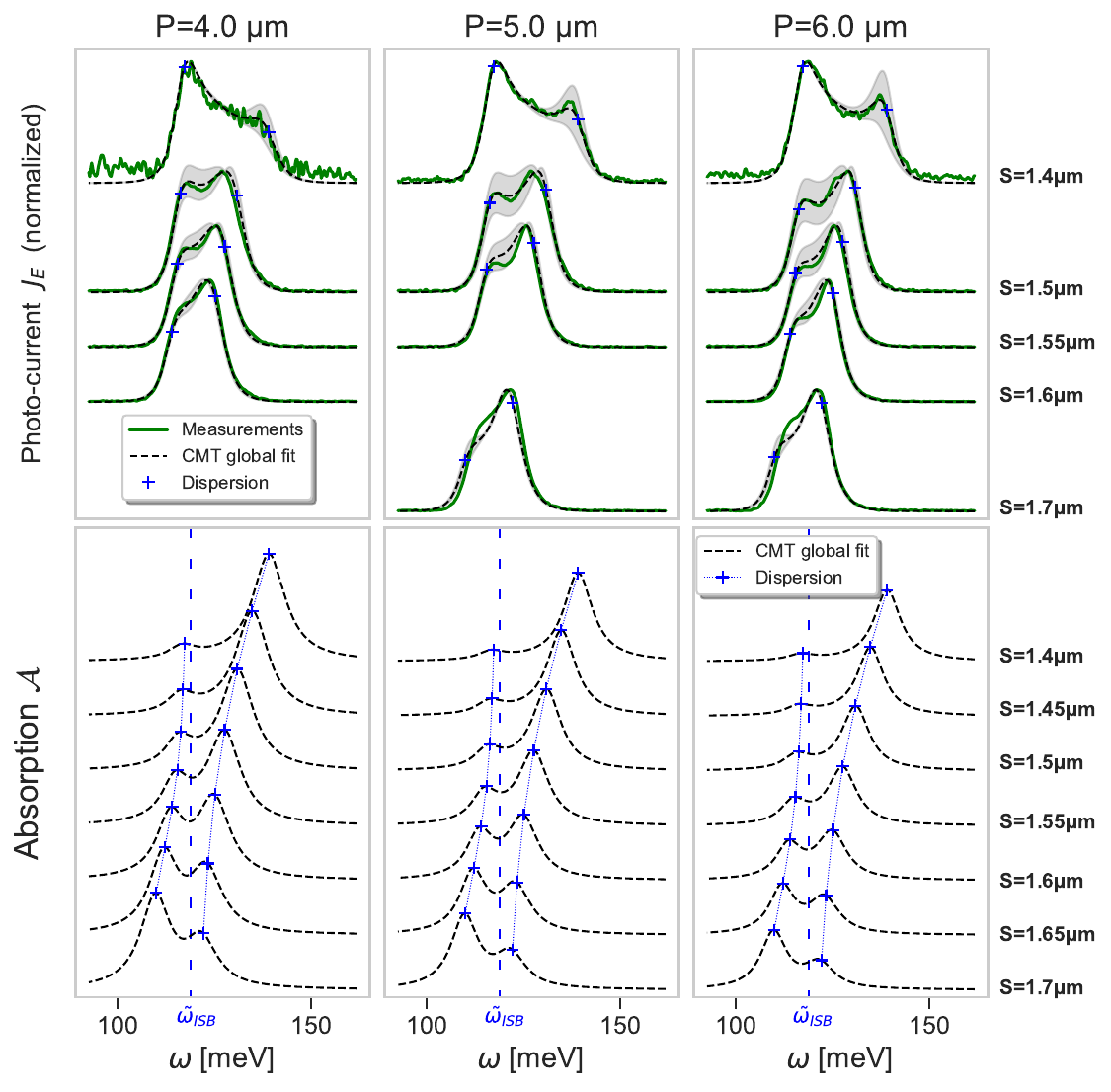} 
\caption{\textit{Upper panel}: additional photo-current measurements (green continuous lines) and \textbf{three resonators} CMT global fit (dashed lines), for different periods $P$ and patch sizes $S$.  Grey areas represent the fit errors propagated on the spectra (see Section I for detailed computation). \textit{Lower pannel}: computed total absorption $\mathcal{A}_\text{tot}$ using Eq.(13) and the parameters extracted from the photo-current fit. For both panels, offsets are added for visibility. and the polaritonic dispersion (blue crosses) is computed considering only the cavity-ISB interaction in the Hamiltonian $\mathcal{H}$ of Eq.(11) of the main text, unperturbed by the extraction mode }\label{supplementary2}
\end{figure*}
\begin{figure*}[ht!]
\centering\includegraphics[width=0.8\textwidth]{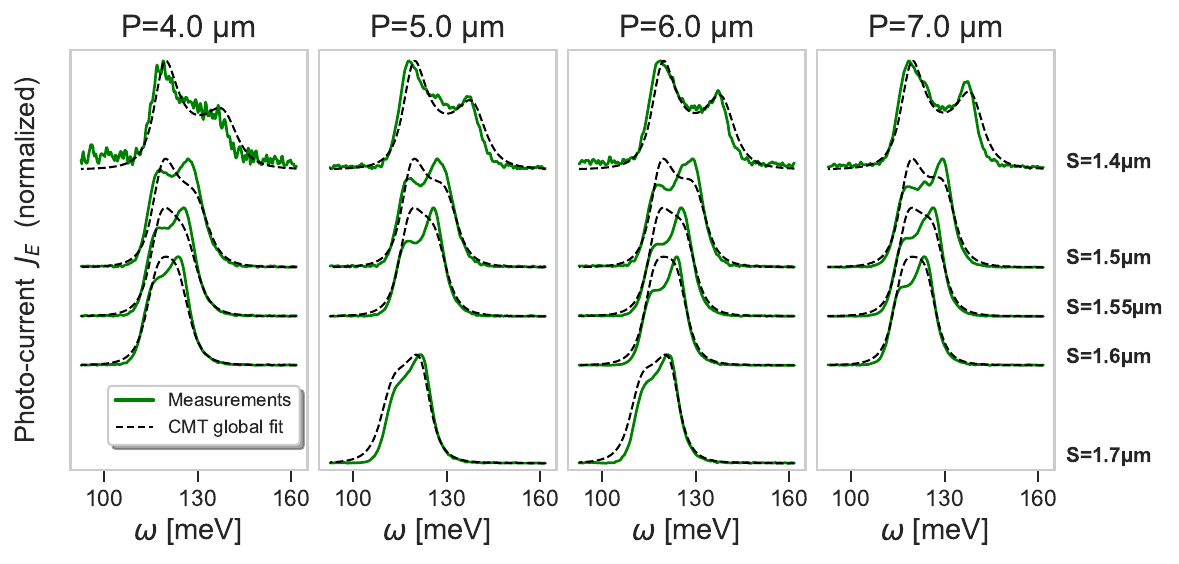}
\caption{Normalized photo-current measurements (green continuous lines) and \textbf{two resonators} CMT global fit (black dashed lines), for different periods $P$ and different patch sizes $S$. Fit procedure is described in Section III C. of the main text. Offsets are added for visibility.} \label{supplementary3}
\end{figure*}